\documentclass[sigconf,nonacm]{acmart}

\usepackage{tikz}
\usepackage{pgfplots}
\usepackage{subcaption}
\usepackage{adjustbox}
\usepackage{booktabs}
\usepackage{multirow}
\usepackage{longtable}
\usepackage{array}
\usepackage{pifont}
\usepackage{draftwatermark}
\SetWatermarkText{Preprint}
\SetWatermarkScale{1}

\AtBeginDocument{%
  }

\begin{document}

\title{LLM Agent-Based Simulation of Student Activities and
Mental Health Using Smartphone Sensing Data}

\author{Wayupuk Sommuang\textsuperscript{*}}
\orcid{0009-0003-6635-6771}
\affiliation{%
  \department{Sirindhorn International Institute of Technology (SIIT)}
  \institution{Thammasat University}
  \city{Pathum Thani}
  \country{Thailand}
}
\email{wayupuk.som@dome.tu.ac.th}

\author{Kun Kerdthaisong\textsuperscript{*}}
\orcid{0009-0005-1168-4983}
\affiliation{%
  \department{Faculty of Engineering}
  \institution{Thammasat School of Engineering, Thammasat University}
  \city{Pathum Thani}
  \country{Thailand}
}
\email{kun.ker@dome.tu.ac.th}

\author{Pasin Buakhaw\textsuperscript{*}}
\orcid{0009-0004-5438-2558}
\affiliation{%
  \institution{Faculty of Engineering, Chulalongkorn University}
  \city{Bangkok}
  \country{Thailand}
}
\email{6833184221@student.chula.ac.th}

\author{Aslan B. Wong}
\orcid{0000-0003-4075-1485}
\affiliation{%
  \department{Artificial Intelligence Research Group}
  \institution{National Electronic and Computer Technology Center (NECTEC)}
  \city{Pathum Thani}
  \country{Thailand}
}
\email{aslan.but@ncr.nstda.or.th}

\author{Nutchanon Yongsatianchot\textsuperscript{\dag}}
\orcid{0000-0003-1332-0727}
\affiliation{%
  \department{Faculty of Engineering}
  \institution{Thammasat School of Engineering, Thammasat University}
  \city{Pathum Thani}
  \country{Thailand}
}
\email{ynutchan@engr.tu.ac.th}

\begin{teaserfigure} 
    \centering
    \includegraphics[width=0.7\linewidth]{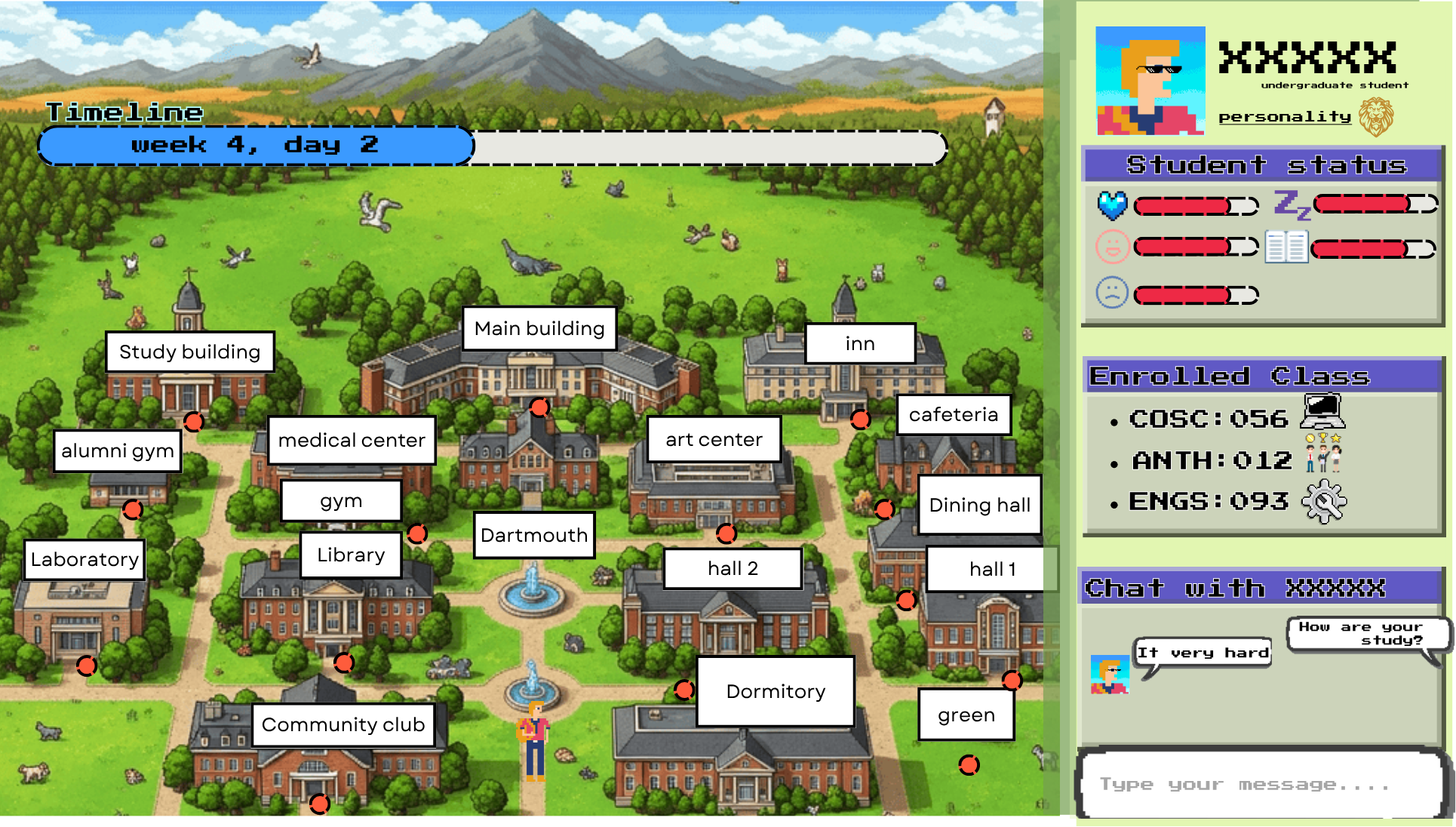}
    \caption{Agent-Based Simulation of Dartmouth College Students}
    \label{fig:figure1}
\end{teaserfigure}

\begin{abstract}
Student mental well-being is vital for academic success, with activities such as studying, socializing, and sleeping playing crucial roles. 
We propose a novel agent-based LLM simulation framework to model student activities and mental health using the StudentLife Dataset. LLM agent was initialized with personality questionnaires and guided by smartphone sensing data throughout a simulated semester. The agent predicts individual behaviors, provides self-reported mental health data through ecological momentary assessments (EMA). 
To ensure accuracy, we investigated multiple LLMs, various prompting techniques, and activity-based mental state management strategies that dynamically update an agent's mental state based on their daily activities. Our simulation results show that GPT-4o-mini outperforms Gemini-2.5-flash in predicting students' stress, sleep, and social levels. The simulation framework allows us to study novel scenarios and students beyond existing datasets, enabling exploration of various experimental manipulations. This study demonstrates the potential of LLM-driven behavioral modeling with sensing data, opening new avenues to understand and support student mental health.
\end{abstract}

\thanks{%
\textsuperscript{*}Equal contribution.
\textsuperscript{\dag}Corresponding author.\\
\url{https://github.com/DarthMouthStudentSimulator/DarthMouthStudentSimulator}
}


\maketitle
\section{Introduction}
Students' mental well-being is a critical factor in college success, yet many students face significant psychological challenges. 
Surveys reveal that a substantial fraction of students struggle with stress and depression; for example, one large study found that approximately 9.5\% of first-year undergraduates were "frequently depressed" and over a third were overwhelmed by academic pressures~\cite{Wang2022FirstGenLens}. To better understand the complex relationship between student well-being, activities, and academic performance, researchers have turned to mobile-sensing technologies. 
The Dartmouth StudentLife study~\cite{studentlife_Wang} pioneered an approach by using smartphones to continuously monitor students' behaviors over a 10-week term, revealing clear temporal patterns: students started strong with positive moods and healthy behaviors, but as academic demands intensified, stress increased while sleep, social interactions, and physical activity declined. 
While mobile sensing data provide rich insights into student behaviour and its correlation with mental health outcomes, agent-based modeling offers a complementary approach that could enrich our understanding by simulating how students experience and navigate a semester. This potentially enables the exploration of intervention strategies and hypothetical scenarios beyond the observed data.

Researchers have shown that LLMs can transform computational social science by simulating human responses and behaviors~\cite{ziems2024can, binz2025foundation}, support psychological research through new measurement and experimental paradigms~\cite{demszky2023using}, and create generative agents that exhibit realistic human-like behavior in simulated environments~\cite{park2023generative}. 
These LLM agents can maintain internal states, communicate naturally, adaptively modify their actions 
(e.g., adjusting schedules or seeking help), and interact with simulated environments~\cite{Gao2024}. 
Through advanced prompting, Frisch and Giulianelli~\cite{frisch-giulianelli-2024-llm} show that agents conditioned on distinct personality profiles exhibit varying degrees of personality consistency and linguistic alignment.

In this study, we leveraged these capabilities to build an LLM agent-based simulation framework for modeling college students, grounded in the Dartmouth StudentLife sensing data, as shown in \autoref{fig:figure1}. Each simulated agent began with a personalized profile, including the Big Five personality traits and baseline questionnaires. The agent maintains its own status, including stamina, sleep performance, stress, mood, and knowledge. As the semester progresses in the Simulation, agents are guided by real smartphone sensing signals (e.g., location, phone use, and social interaction patterns). Using carefully designed prompt templates, we enabled agents to update their mental states based on their activities, complete assignments, and take exams. This approach allows us to explore how LLM agents respond to sensing data, activity patterns, and mental states, providing a new perspective for studying the dynamics of student well-being.

Our contributions are as follows. First, we introduce a novel LLM agent-based simulation framework specifically designed to model student activities and mental well-being by leveraging real-world smartphone sensing data from the StudentLife Dataset. Second, we demonstrate how this framework can replicate the observed behavioral and mental health patterns. Finally, the framework paves the way for studying the complex interplay between student behaviours and their psychological states by simulating and analysing the factors influencing student success and mental health.

\section{Related Work}

\subsection{Mobile sensing and StudentLife dataset}

Mobile sensing has emerged as a powerful tool for passively collecting rich longitudinal data on human behavior and well-being in ecological environments. 
Previous research has extensively used the StudentLife dataset to explore the complex relationships between student behavior and mental health. For example, StudentLife ~\cite{studentlife_Wang} demonstrated correlations between sleep patterns, social interactions, and academic performance with reported levels of stress and depression. Subsequent studies have used various statistical analyses and traditional machine learning models to predict mental health states ~\cite{nepal2025predicting}, identify behavioral markers of stress ~\cite{harari2016smartphones}, and understand the impact of academic workload on well-being ~\cite{nepal2025predicting}. Beyond StudentLife, other studies have also leveraged mobile sensing for similar purposes, such as understanding daily routines and their impact on mood ~\cite{li2012daily} or detecting symptoms of depression through passive phone data ~\cite{saeb2015relationship}. These efforts highlight the utility of ubiquitous sensing for capturing the intricate dynamics of student life and mental health in a non-intrusive manner, thereby laying the groundwork for data-driven interventions.

\subsection{LLM Agent-Based Simulation}

Recent studies have demonstrated the potential of LLMs to create realistic agent-based simulations. 
Park et al.~\cite{park2023generative} introduced generative agents that can simulate believable human behavior in interactive environments, maintain memories, and plan actions based on their experiences. Although their agents exhibited complex social behaviors, they operated in fully synthetic environments without grounding in real-world behavioral data. In educational contexts, researchers have explored LLM-based student modeling. Lu et al.\cite{generative_student_lu} proposed Generative Students that use LLMs to simulate student profiles and generate responses to multiple-choice questions based on the Knowledge-Learning-Instruction framework. Similarly, EduAgent\cite{eduagent_xu} models fine-grained student learning behaviors by integrating large language models with cognitive science. 
However, the approaches primarily emphasize academic performance simulation, with limited attention to real-world behavioral patterns and mental health dynamics that significantly shape students' lived experiences.

\subsection{LLMs for Personality and Psychological Assessment}

Parallel research has explored the use of LLMs to model and assess personality traits. 
Dong et al.\cite{Humanizing_llm_dong} systematically reviewed the application of psychological theories to LLMs, demonstrating that LLMs can exhibit reproducible personality patterns under specific prompting schemes. Zhu et al.~\cite{zhu2025investigating} showed that LLMs could infer the Big Five personality traits from conversations using zero-shot prompting, with improved accuracy when incorporating intermediate steps, such as first inferring BFI-10 item scores before calculating traits. While these studies demonstrate LLMs' capability  to model psychological constructs, they primarily focus on static personality assessment from text data rather than dynamic behavioral simulation guided by real-world activity patterns. Our work bridges this gap by grounding LLM agents in continuous mobile sensing data, enabling the simulation of how student behaviors, activities, and mental states evolve throughout a semester based on actual behavioral traces. This approach uniquely combines the psychological modeling capabilities of LLMs with rich behavioral insights from ubiquitous computing, opening new possibilities for understanding and supporting student well-being.

\begin{figure*}
    \centering
    \includegraphics[width=0.85\linewidth]{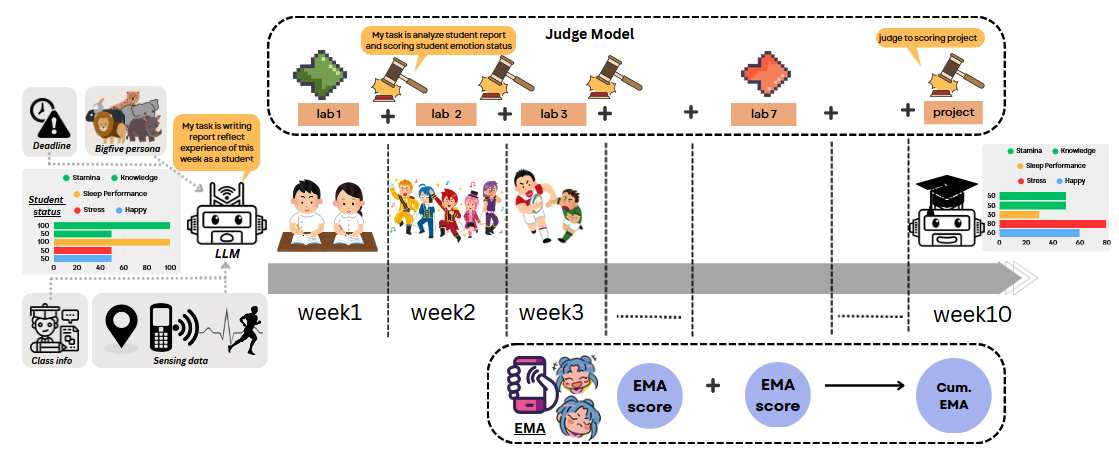}
    \caption{LLM Agent-Based Simulation framework. Each week students will receive lab assessments and answer EMA. In the last week, the judge will calculate cumulative score and give a score of final project.}
    \label{fig:Framework}
\end{figure*}

\section{Simulation Framework}
\label{Simulation Framework}

\autoref{fig:Framework} illustrates our LLM agent-based simulation framework designed to capture student experiences. We begin by assigning each agent a \textbf{student profile}(Section 3.1), including personality, and a dynamic \textbf{student status} that adapts to life events. The Simulation progressed weekly. Each week, the student agent submits a \textbf{self-reflection report} to a \textbf{judge agent} (Section 3.3), which analyzes the content and updates the student's mental state. Finally, the student agent completes an \textbf{Ecological Momentary Assessment (EMA)} and academic tasks (assignments or exams) to facilitate the evaluation and generate a weekly summary. 

\subsection{Students}
\label{Student}

The objective of the student agent is to simulate the behavior of 26 undergraduate students at Dartmouth College who enrolled in the COSC 065 Smartphone Programming in the StudentLife Dataset. To achieve this, we constructed a detailed student profile that includes multiple components, such as \textit{ Big Five personality, various mental state statuses, enrolled courses, sensing data, and academic records }, as shown in Figure \ref{fig:Framework}.

\subsubsection{\textbf{Student's Status}}
\label{sec:Student's Status}

The student's status consists of multiple dimensions to capture the dynamic internal state of a student agent. Each dimension is quantified and updated over time to reflect an agent's response to life events and academic situations. Together, these dimensions form a comprehensive student status that is updated weekly, allowing the system to simulate and track realistic psychological changes over time. The Dimensions are as follows:

\begin{itemize}
    \item \textbf{Mood:}  
    A set of emotional indicators such as \textit{stress} and \textit{happiness} were each scored on a scale of 0 to 100. 
    \item \textbf{Stamina}  
    indicates the agent's physical and mental energy.
    \item \textbf{Knowledge}  
    represents the agent's perceived academic competence.
    \item \textbf{Sleep Performance}  
    measured the quality and duration of sleep.
    
\end{itemize}

\subsubsection{\textbf{Sensing data}}
\label{sec:Sensing data}
We used aligned activity and GPS data with timestamps for the sensing data. To help student agents understand their activities, we converted latitude and longitude to location and description of location in text format and provided them with a weekly routine report.
Furthermore, we split the data into \textit{n} weeks with 7 days and 24 h, and if there were no data at that timestamp, we filled it with \textit{null}.

\subsection{\textbf{Student Lab Assessment}} 
\label{sec:Student Lab Assessment}
In lab assessment, student agent must complete multiple-choice test for each week (week 2 to week 7) based on \cite{campbell_smartphone_programming_2013}, 10 question a week for six topics consist of \textit{1. Layouts \& Views Basics}, \textit{2. UI Components \& Event Handling}, \textit{3. Activities  and Intents}, \textit{4. Layouts \& UI Design}, \textit{5. ListView \& ArrayAdapter}, \textit{6. Data Storage} total 70 marks generated by GPT-4o~\cite{openai_gpt4o_2025}. 
See timeline in Figure~\ref{fig:Framework}. 
Finally, at the end of the semester (week 10), the student agents submitted the project report which was graded by the LLM judge. 

\subsection{Judge}
\label{Judge}

We prompted the LLM-based model,to perform mental state analysis using the students' weekly self-reflection reports. All prompts and students' academic performances are provided in the Supplementary Material. This analysis was used to update their mental state and provide reasons for students experiences after completing their weekly routines, as shown in Framework at Figure~\ref{fig:Framework} 

\begin{figure}
\centering
\includegraphics[width=1.0\linewidth]{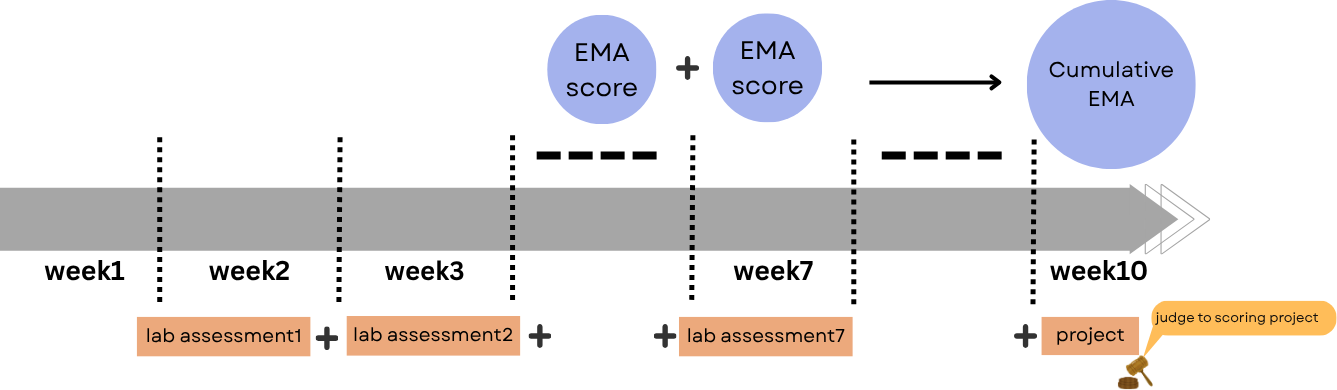}
   \caption{Timeline: Each week students will receive lab assessments and answer EMA. In the last week, the judge will calculate cumulative score and give a score of final project.}
   \label{fig:timeline}
\end{figure}

\section{Preliminary Results} 
\label{Preliminary Results}

In this study, we evaluated the performance of two prominent LLMs, GPT-4o-mini and Gemini-2.5-Flash, in predicting student well-being metrics based on simulated data. \autoref{table:model_performance} summarizes the mean absolute error (MAE) and root mean squared error (RMSE) between the LLM-predicted student status and the ground truth EMA responses--stress, sleep, and social (based on noise data) levels. Our analysis yielded several key observations as follows. First, GPT-4o-mini consistently outperformed Gemini-2.5-Flash across all evaluated metrics. This superior performance suggests that GPT-4o-mini may exhibit a stronger capacity for aligning with human-like behavioural and psychological state modeling, which is crucial for the fidelity of agent-based simulation.

Second, the prediction of the social level achieved the highest performance for both models (MAE: 0.250 for GPT-4o-mini, 0.282 for Gemini-2.5-Flash). We hypothesised that this outcome was due to the inherent clarity and consistency with which the social dimensions of the Big Five personality traits (e.g. introversion vs. extraversion) manifest in observable social behaviours. These traits likely provide a more stable and direct basis for prediction than other more volatile student states. However, neither model demonstrated high accuracy in predicting stress and sleep levels. This limitation may stem from the fact that stress and sleep are often profoundly influenced by transient situational and contextual factors (e.g. upcoming exams, unexpected events, environmental noise) that are not fully captured or consistently predictable from stable personality profiles. Future studies should explore the incorporation of real-time contextual data to improve these predictions.

\begin{figure}
    \centering
    \includegraphics[width=1\linewidth]{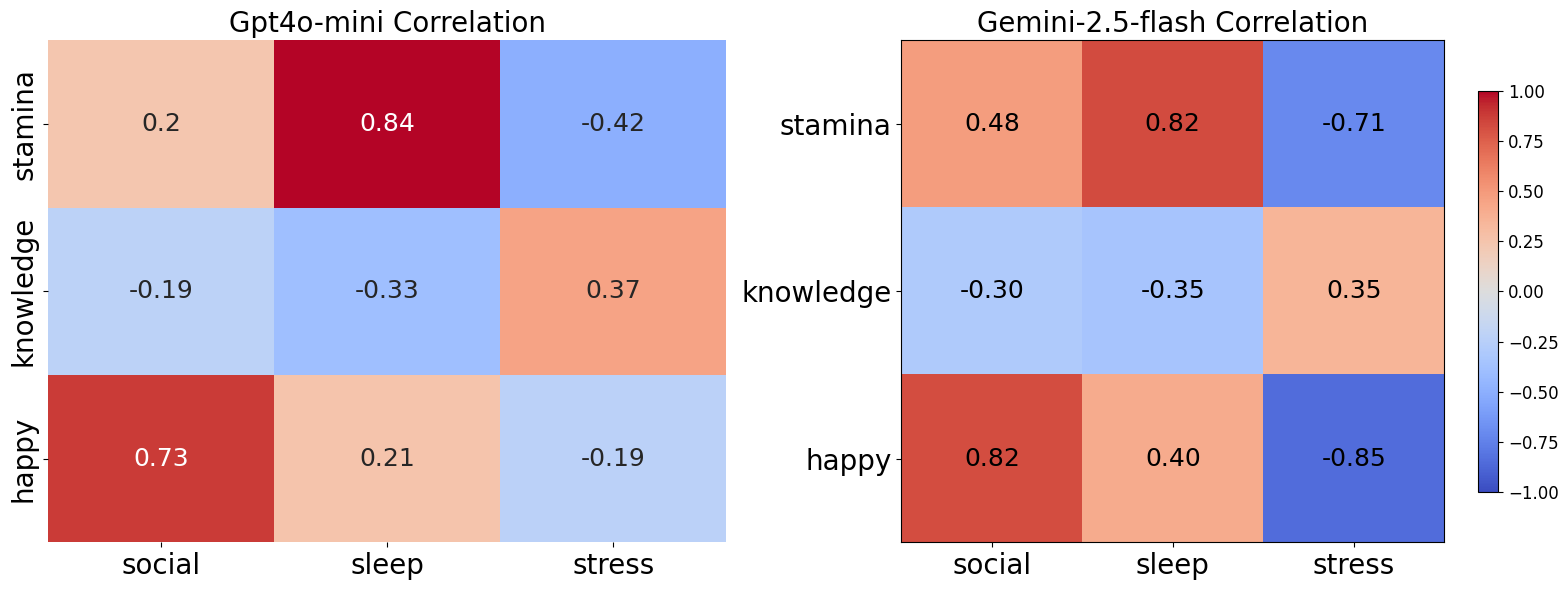}
    \caption{Spearman correlation on social, sleep, and stress}
    \label{fig:two_correlation}
\end{figure}

Furthermore, we analyze the relationships between the predicted EMA values and other simulated student status dimensions (happy, knowledge, and stamina) using Spearman correlation, as illustrated in \autoref{fig:two_correlation}. We found that student agents tend to report higher happiness when they experience greater social engagement and adequate rest. Stamina is strongly correlated with social activity and sleep performance, whereas all three show a negative relationship with stress. Interestingly, knowledge often increases alongside stress, but there is a slight negative correlation with social interaction and sleep, highlighting the potential trade-offs between academic intensity and other well-being factors.

\begin{table}
\centering
\setlength{\tabcolsep}{1.5mm}
\begin{tabular}{lcccccc}
\toprule
\multirow{2}{*}{\textbf{Status}} & \multicolumn{2}{c}{\textbf{Gemini-2.5-flash}\cite{google2025gemini2.5}} & \multicolumn{2}{c}{\textbf{GPT-4o-mini}\cite{openai_gpt4o_mini}} \\
 & MAE & RMSE & MAE & RMSE \\
\midrule
Stress level & 0.750 & 0.873 & 0.675 & 0.795 \\
Sleep level  & 1.245 & 1.329 & 0.963 & 1.080 \\
Social level & 0.282 & 0.329 & 0.250 & 0.274 \\
\bottomrule
\end{tabular}
\caption{Comparison of model performance (MAE and RMSE) on predicting cumulative EMA (stress, sleep, social levels) of 26 participants.}
\label{table:model_performance}
\end{table}

\section{Conclusion and Future Works}
\label{Discussion}

This study introduces our LLM agent-based simulation framework for modeling student behaviour and mental health and presents initial findings on the feasibility of guiding LLM-powered student agents using mobile sensing data. While promising, 
There is still significant work ahead to fully unlock the potential of this method. 
We plan to investigate new ways to evaluate academic performance beyond the current metrics and to better understand how students' mental states influence their performance. Finally, we will implement a memory system to ensure that the agents maintain coherent states throughout the entire simulated semester.

Our framework can go beyond simply replicating existing StudentLife data and allows us to explore novel scenarios and new student profiles. 
We plan to expand the simulation to include multiple interacting agents, enabling us to study complex dynamics, such as social influence, peer effects on mental health, and the impact of social media through agent-to-agent interactions. 
The Simulation also opens doors for intervention studies by manipulating parameters such as activity patterns, personality traits, and the classroom structure. These experiments could help pinpoint the behavioral changes that genuinely improve student well-being. Moreover, these LLM agents can engage in hypothetical interviews, offering rich qualitative insights into their simulated mental health experiences, which conventional models cannot provide. Ultimately, our work highlights the immense promise of combining LLM capabilities with mobile sensing data for behavioral modeling purposes. This integration creates new avenues for understanding and addressing student mental health challenges. By bridging LLM-based agents with real-world behavioral data, we open up unprecedented opportunities to grasp the complex dynamics of student well-being.

\begin{acks}
    
We  would like to express our sincere gratitude to  PreceptorAI and CARIVA Thailand for providing computing resources. The authors also acknowledge the Faculty of Engineering, Thammasat School of Engineering, Thammasat University, for their support.    
\end{acks}

\appendix

\section{Dataset Details}
\label{dataset-details}

\subsubsection{Agent Initialization}
To calculate BigFive personality score we map student questionnaire result into the big five personality test then calculate the personality score 

\subsubsection{Data Anonymization and Privacy}
All student data underwent comprehensive anonymization to protect participant privacy. Personal identifiers were removed and replaced with anonymous UIDs (u01-u25). Location data was generalized to campus zones rather than precise coordinates. Timestamps were normalized to relative weeks (Week 1-10) instead of absolute dates. Demographic information was aggregated to prevent individual identification.

\subsubsection{Dataset Limitations}
\begin{itemize}
\item \textbf{Temporal Scope}: Limited to 10-week period
\item \textbf{Institutional Bias}: Single university setting
\item \textbf{Demographic Homogeneity}: Predominantly undergraduate students
\item \textbf{EMA ground truth}: unstable(depends on student) and non-periodically.
\end{itemize}

\subsection{Additional Results}
\label{Addtional results}

\begin{figure}
    \centering
    \includegraphics[width=1.0\linewidth]{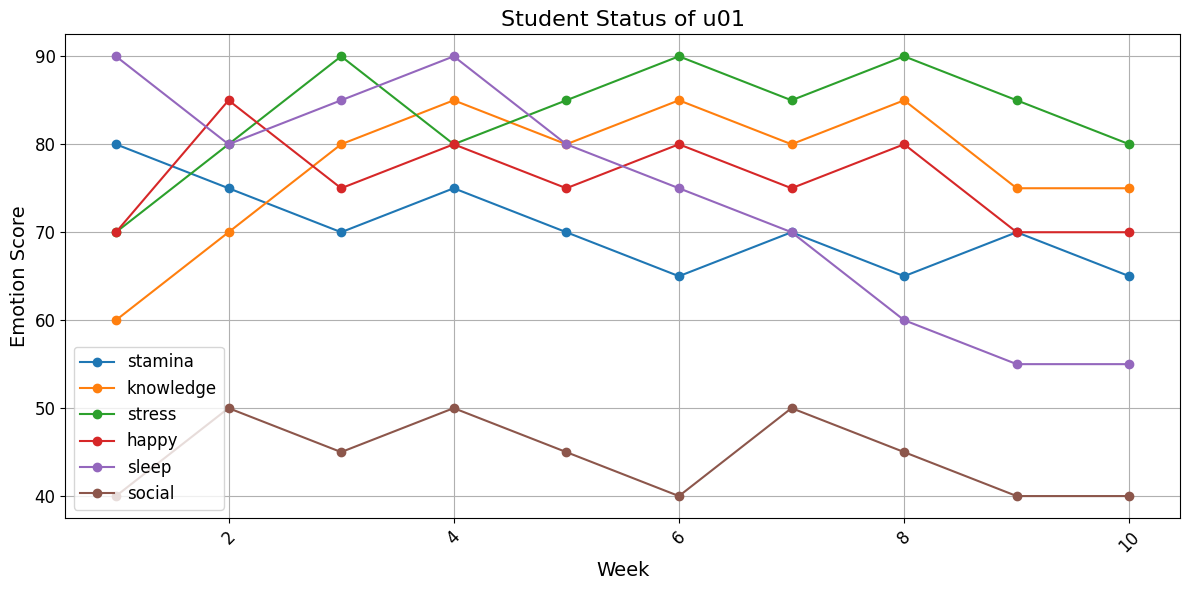}
    \caption{Student status timelines per 10 week from student agent u01 predicted by Gpt4o-mini}
    \label{fig:result_u01}
\end{figure}

\begin{figure}
    \centering
    \includegraphics[width=0.5\linewidth]{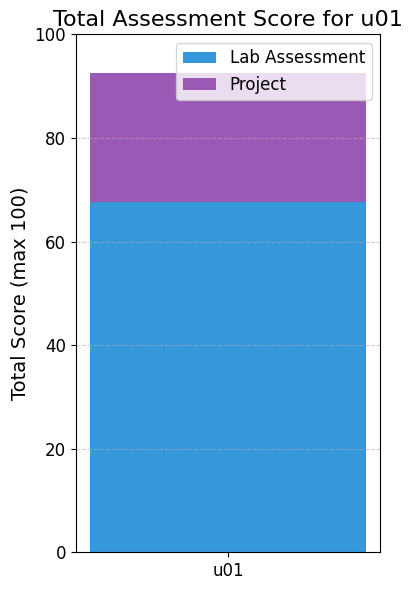}
    \caption{total score of u01 in smartphone programming predicted by Gpt4o-mini}
    \label{fig:u01_total_score}
\end{figure}

\begin{figure}
    \centering
    \includegraphics[width=1.0\linewidth]{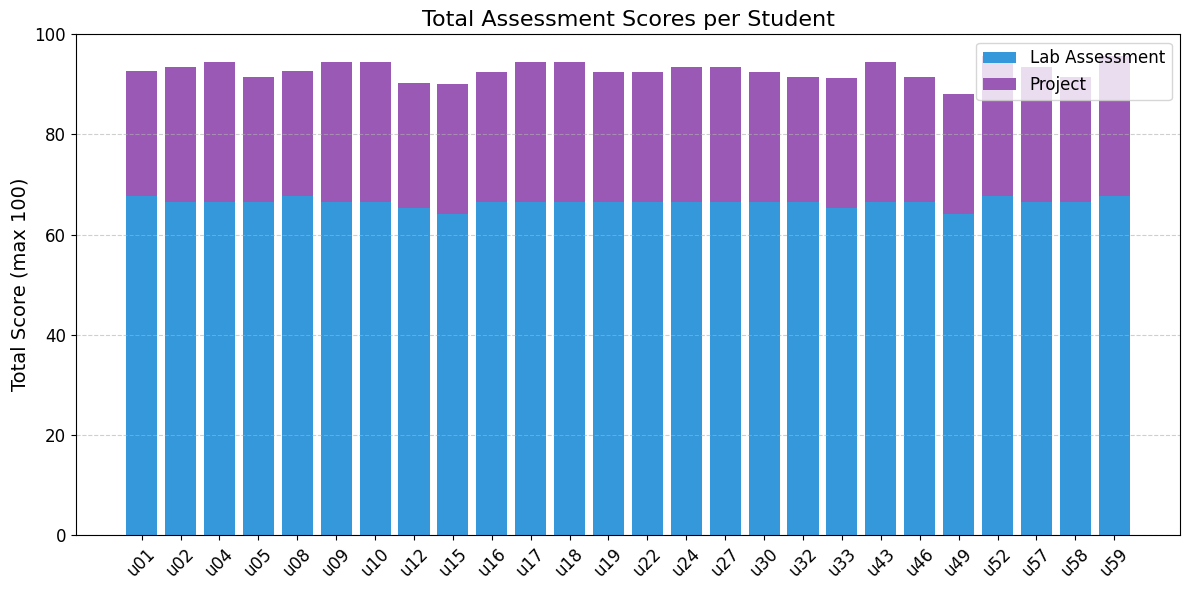}
    \caption{total score of all student who enrolled smartphone programming predicted by Gpt4o-mini}
    \label{fig:total_score_all_students}
\end{figure}

\section{Project Assessment}
\label{Project Assessment}

At week 10 (end of the semester), we used an LLM-based model, the same as the student agent, to judge the score of their project, in a total of 30 points across the following \textbf{criteria}:
\begin{itemize}
    \item Is the idea innovative or unique compared to existing apps?
    \item Does it clearly address a real problem or user need?
    \item  Is the scope appropriate (not too simple, not too ambitious)?
    \item  Does it demonstrate thoughtful consideration of user experience and impact?
\end{itemize}.

\section{Prompts}
\label{Prompts}

\subsection{Student Agent}
\label{Student Agent}

\subsubsection{Journal Entry Generation}
\label{subsec:journal-prompt}

The student agent generates weekly self-reflection journal entries based on personality traits, class schedule, and sensing data. The system prompt establishes the agent's persona and context:

\begin{quote}
\textit{System Prompt:}
You are a university student simulator. You will generate a self-reflection journal based on class schedule and real-world sensing data. Write naturally and personally, as if you were the student reflecting on your week. Focus only on context - DO NOT add unnecessary elements like name or date.

Personality:
- Openness: \{O\_score\}
- Conscientiousness: \{C\_score\}
- Extraversion: \{E\_score\}
- Agreeableness: \{A\_score\}
- Neuroticism: \{N\_score\}

Enrolled Classes:
\{formatted\_class\_schedule\}

Current Student status: 
- happy: \{emotion\_status.happy\}
- sleep: \{emotion\_status.sleep\}
- social: \{emotion\_status.social\}
- stamina: \{emotion\_status.stamina\}
- stress: \{emotion\_status.stress\}
- knowledge: \{emotion\_status.knowledge\}

Your Class Experience Summary:
\{class\_experience\_summary\}
\end{quote}

\begin{quote}
\textit{User Prompt:}
You are a university student. This is your weekly activity.
Sensing Data for Week:
(Each entry: Timestamp | Activity | Location | Location description)
\{sensing\_data\_formatted\}

TASK: Reflect on your experience this week in class, on campus, and in your social life. How did you feel? Any challenges? What are your goals for next week?
\end{quote}

\subsubsection{Project Submission Generation}
\label{subsec:project-prompt}

For generating creative project ideas, the student agent uses the following prompts:

\begin{quote}
\textit{System Prompt:}
You are a university student simulator.
Personality:
- Openness: \{O\_score\}
- Conscientiousness: \{C\_score\}
- Extraversion: \{E\_score\}
- Agreeableness: \{A\_score\}
- Neuroticism: \{N\_score\}

Enrolled Classes:
\{formatted\_class\_schedule\}

Current Student status: 
- happy: \{emotion\_status.happy\}
- sleep: \{emotion\_status.sleep\}
- social: \{emotion\_status.social\}
- stamina: \{emotion\_status.stamina\}
- stress: \{emotion\_status.stress\}
- knowledge: \{emotion\_status.knowledge\}
\end{quote}

\begin{quote}
\textit{User Prompt:}
You are a university student. This is your last week to present final project(ideas) on smartphone programming to get 30 score.

Please generate a creative and feasible mobile app project idea that demonstrates your understanding of smartphone programming concepts.
\end{quote}

\subsubsection{Emotion Analysis}
\label{subsec:emotion-prompt}

The student agent analyzes emotional states from journal entries using:

\begin{quote}
\textit{System Prompt:}
You are an emotional state analyzer. Your task is to analyze a student's weekly self-reflection journal and infer their emotional state.

You must:
1. Output a Python dictionary with keys: ['stamina', 'knowledge', 'stress', 'happy', 'sleep', 'social']
- Each value should be an integer between 0 and 100.

2. Explain briefly why each emotional value was chosen.
- Use reasoning directly from the journal text.
- Match student words/phrases with your judgment.

Current Student status: 
\{current\_emotion\_status\}

Output format:
\{
"stamina": value,
"knowledge": value,
"stress": value,
"happy": value,
"sleep": value,
"social": value
\}

Reasoning:
- Stamina: because the student mentioned feeling drained after class.
- Stress: because they worried about deadlines, etc.
\end{quote}

\begin{quote}
\textit{User Prompt:}
Here is the journal entry from the student:

\{journal\_text\}

Please analyze and output both the emotional dictionary and reasoning.
\end{quote}

\subsection{Academic Evaluator}
\label{subsec:evaluator}

\subsubsection{Weekly Exam Evaluation}
\label{subsec:exam-prompt}

The academic evaluator simulates student performance on weekly exams using personality and emotional state:

\begin{quote}
\textit{Exam Prompt:}
You are a student with the following characteristics:
- Big Five Personality: Openness=\{O\_score\}, Conscientiousness=\{C\_score\}, Extraversion=\{E\_score\}, Agreeableness=\{A\_score\}, Neuroticism=\{N\_score\}
- Current Status: Stamina=\{stamina\}, Knowledge=\{knowledge\}, Stress=\{stress\}, Happy=\{happy\}, Sleep=\{sleep\}, Social=\{social\}

You are taking a smartphone programming class exam. Here's the question:

Topic: \{topic\}
Question: \{question\}

Please provide your answer as a single letter (A, B, C, or D).
\end{quote}

\subsubsection{Project Scoring}
\label{subsec:project-scoring}

The academic evaluator assesses project submissions using:

\begin{quote}
\textit{System Prompt:}
You are an expert university instructor and judge for a smartphone programming class. Your task is to evaluate student mobile app project ideas based on the following criteria.

Evaluation Criteria (30 - Project Idea):
- Is the idea innovative or unique compared to existing apps?
- Does it clearly address a real problem or user need?
- Is the idea technically feasible for implementation by a student team within one semester?
- Is the scope appropriate (not too simple, not too ambitious)?
- Does it demonstrate thoughtful consideration of user experience and impact?

Instructions:
1. Evaluate the idea out of 30 based on the above criteria.
2. Answer in form of x/30
Remember answer in number/30
\end{quote}

\begin{quote}
\textit{User Prompt:}
Student Submission:
\{submission\_text\}

Please provide your evaluation.
\end{quote}
\end{document}